\documentclass[conference]{IEEEtran}
\IEEEoverridecommandlockouts
\usepackage{cite}
\usepackage{amsmath,amssymb,amsfonts}
\usepackage{algorithmic}
\usepackage{graphicx}
\usepackage{textcomp}
\usepackage{xcolor}
\def\BibTeX{{\rm B\kern-.05em{\sc i\kern-.025em b}\kern-.08em
    T\kern-.1667em\lower.7ex\hbox{E}\kern-.125emX}}
\begin{document}

\title{Accuracy Simulation of MF R-Mode Systems Using TOA Variance
\thanks{This work was supported in part by Grant RS-2024-00407003 from the ``Development of Advanced Technology for Terrestrial Radionavigation System'' project, funded by the Ministry of Oceans and Fisheries, Republic of Korea;
in part by the National Research Foundation of Korea (NRF), funded by the Korean government (Ministry of Science and ICT), under Grant RS-2024-00358298; 
in part by the Future Space Navigation and Satellite Research Center through the NRF, funded by the Ministry of Science and ICT (MSIT), Republic of Korea, under Grant 2022M1A3C2074404; 
and in part by the MSIT, Korea, under the Information Technology Research Center (ITRC) support program supervised by the Institute of Information \& Communications Technology Planning \& Evaluation (IITP) under Grant IITP-2024-RS-2024-00437494.
}
}

\author{\IEEEauthorblockN{Jaewon Yu} 
\IEEEauthorblockA{\textit{School of Integrated Technology} \\
\textit{Yonsei University}\\
Incheon, Korea \\
jaewon.yu@yonsei.ac.kr} 
\and
\IEEEauthorblockN{Pyo-Woong Son${}^{*}$} 
\IEEEauthorblockA{\textit{Department of Electronics Engineering} \\
\textit{Chungbuk National University} \\
Cheongju, Korea \\
pwson@cbnu.ac.kr}
{\small${}^{*}$ Corresponding author}
}

\maketitle

\begin{abstract}
To ensure consistent navigation services despite GNSS signal disruptions, Korea is developing the R-Mode system. 
This study focuses on enhancing the simulation accuracy of the MF R-Mode system's performance by integrating data from the Eocheong transmitter with existing data from the Palmi and Chungju transmitters. 
Additional measurements from these three transmitters were gathered using the DARBS receiver to model the Time-of-Arrival (TOA) variance. 
Analysis of this data facilitated the calculation of new constants and transmitter-specific jitter values, which were then used to determine coverage areas based on the updated parameters.
\end{abstract}


\section{Introduction}
Global Navigation Satellite Systems (GNSS) \cite{Kim14:Comprehensive, Chen11:Real, Lee23:Seamless, Kim23:Machine, Kim23:Low, Lee24:A, Kim23:Single, Kim22:Machine, Lee22:Urban} provide highly accurate positioning, navigation, and timing (PNT) information by receiving signals transmitted from satellites. 
However, these signals are vulnerable to radio frequency interference (RFI) \cite{Park21:Single, Park18:Dual, Kim19:Mitigation, Park17:Adaptive, Jeong20:RSS, Moon24:HELPS, Lee22:Performance} and ionospheric anomalies \cite{Jiao15:Comparison, Sun20:Performance,  Seo11:Availability, Lee22:Optimal, Sun21:Markov, Lee17:Monitoring, Ahmed17:Statistical}. 
North Korea has conducted several GPS jamming attacks, impacting various sectors including maritime, communication, and aviation in South Korea. \cite{Kim22:First, Rhee21:Enhanced, Son20:eLoran, Son24:eLoran}.

To ensure stable and independent navigation services as a backup to GPS, South Korea is considering the eLoran system\cite{Son23:Demonstration, Kim22:First, Son22:Compensation}. 
Based on the Loran-C system, eLoran provides PNT services by transmitting powerful signals from multiple ground-based transmitters. 
From 2016 to 2020, South Korea developed eLoran technology, establishing a time synchronization system at the existing Pohang and Gwangju transmitters and installing a low-power transmitter at the Incheon test site. 
In 2023, the Incheon test site's low-power transmitter was relocated to Socheong Island in the northern West Sea, increasing the effective radiated power to 8 kW. 
Currently, Korea's eLoran system consists of three main transmitters (Pohang, Gwangju, Socheong Island) and two differential Loran stations (Incheon, Pyeongtaek), providing a foundation for reliable navigation services even in GPS signal interference situations\cite{Son24:eLoran}.

Recently, Korea has sought to reduce the high initial installation costs of eLoran transmitters by upgrading its communication infrastructure\cite{Son22:Analysis, Son22:Development}. 
This effort includes leveraging Medium Frequency (MF) and Very High Frequency (VHF) signals as part of the R-Mode project, which can cover a wider area by utilizing the already existing infrastructure.
Studies have been conducted to determine the parameters necessary for calculating signal strength and to apply these parameters in simulation tools to model the signal strength of the Yeongju transmitter\cite{Yu22:Simulation}.
Additionally, research has been undertaken to model the variance in Time of Arrival (TOA) measurements using data collected from the Palmi and Chungju transmitters\cite{Yu23:Empirical}.
However, these studies were limited to data from only two transmitters, making it difficult to accurately assess the system's overall performance.

In this paper, we enhance the modeling of TOA variance by integrating additional measurements, including data from the Eocheong transmitter. By incorporating these new data, we improve the accuracy of system performance calculations, overcoming previous limitations and resulting in more precise simulation outcomes.
Data were gathered using the DARBS receiver, and the parameters for the Eocheong, Palmi, and Chungju transmitters were re-estimated by integrating these new data into the previous research datasets. 
(More details about the receiver can be found at https://darbs.co.kr/.)
A coverage map created using the combined data from the three transmitters was produced to evaluate system performance.

\section{Methodology}
\subsection{Estimation of Parameters for MF R-Mode TOA Measurements}

The TOA variance formula for the MF R-Mode system follows \cite{Yu23:Empirical}:

\begin{equation}
 \sigma_i^2 = J_i^2 + \frac{C^2}{SNR_i}
 \label{eqn:MF_R-Mode_TOA_variance}
\end{equation}

Jitter refers to random errors originating from the transmitter and is independent of the receiver's location, caused by thermal noise and other reasons\cite{Rhee21:Enhanced, Lo08:Loran}.
$J_i$ represents the jitter of transmitter $i$, and $C$ is a constant value, both of which are parameters estimated from actual measurement data. 
$\sigma$ denotes the standard deviation of the MF R-Mode TOA measurements, while $SNR_i$ represents the signal-to-noise ratio of the signal received from transmitter $i$. 
The constants $C$ and jitter values in  \ref{eqn:MF_R-Mode_TOA_variance} are parameters that can be updated as new data is incorporated.

\subsection{Data Collection and Processing}

In this research, alongside the data obtained from \cite{Yu23:Empirical}, we use raw phase measurements and SNR data collected using the DARBS medium frequency receiver from Palmi, Chungju, and Eocheong in Korea. 
The SNR data is taken from the receiver's measurements, and the formula for calculating $\sigma^2_i$ from the raw phase measurements follows \cite{Yu23:Empirical}:

\begin{equation}
 \sigma^2_i = \text{Var} (\text{TOA}_i) = \left( \frac{\lambda}{2\pi} \right)^2 \cdot \text{Var} (\phi_{\text{cont},i})
 \label{eqn:relation}
\end{equation}

Here, $\lambda$ denotes the wavelength of the CW signal. 
The parameters $J_i$ and $C$ are derived by minimizing the residual sum of squares (RSS) between the model in (\ref{eqn:relation}) and the actual measurements, based on the observed $SNR_i$ and $\sigma^2_i$ values.

\subsection{Accuracy Calculation of MF R-Mode}

The approach for determining the signal accuracy follows the same procedure as outlined in \cite{Rhee21:Enhanced, Lo08:Loran}. All equations in this section are from \cite{Rhee21:Enhanced}.

The geometry matrix $G$, which is detailed in (\ref{eqn:azimuth}), is constructed using the sine and cosine of the angles between the user and each transmitter, $\theta_i$.

\begin{equation}
 G = \begin{bmatrix}
 \cos({\theta}_{1}) & \sin({\theta}_{1}) & 1 \\
 \cos({\theta}_{2}) & \sin({\theta}_{2}) & 1 \\
 \cos({\theta}_{3}) & \sin({\theta}_{3}) & 1 \\
 \end{bmatrix}
 \label{eqn:azimuth}
 \end{equation}

The matrix $R$, which is based on the variances $\sigma_i^2$, is defined as follows:

\begin{equation}
 R = \begin{bmatrix}
 \sigma^2_{\text{1}} & 0 & 0 \\
 0 & \sigma^2_{\text{2}} & 0 \\
 0 & 0 & \sigma^2_{\text{3}} \\
 \end{bmatrix}
 \label{eqn:diagonal} 
 \end{equation}

Using this matrix, the error covariance matrix $K$ for the position estimation is determined by (\ref{eqn:posionerrmat}).

\begin{equation}
 K = (G^T R^{-1} G)^{-1}
 \label{eqn:posionerrmat}
 \end{equation}

The 95\% horizontal accuracy is derived using the error covariance matrix for the position estimation, as shown in (\ref{eqn:accuracy}).

\begin{equation}
 Accuracy = 2\sqrt{K_{11} + K_{22}}
 \label{eqn:accuracy}
 \end{equation}

$K_{11}$ and $K_{22}$ are the diagonal components of the error matrix, indicating the variance in the position estimation.
The accuracy calculation assumes the position solution is unbiased and that biases from ASF are corrected through spatial and temporal ASF adjustments. 
This type of accuracy is known as repeatable accuracy.

\section{Results}

\subsection{Results of MF R-Mode Parameter Estimation}
\begin{figure}
    \centering
    \includegraphics[width=0.8\linewidth]{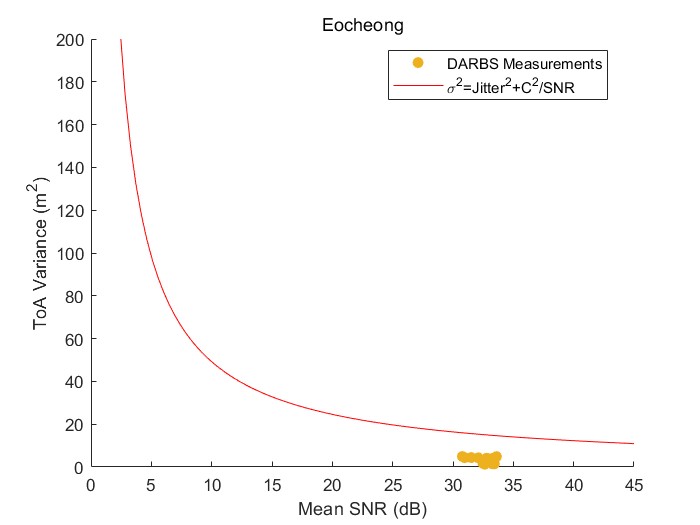}
    \caption{Model fit for Eocheong transmitter. Yellow points represent measurements from the DARBS receiver and the red curve represents (\ref{eqn:MF_R-Mode_TOA_variance}).}
    \label{fig:Eocheong}
\end{figure}

\begin{figure}
    \centering
    \includegraphics[width=0.8\linewidth]{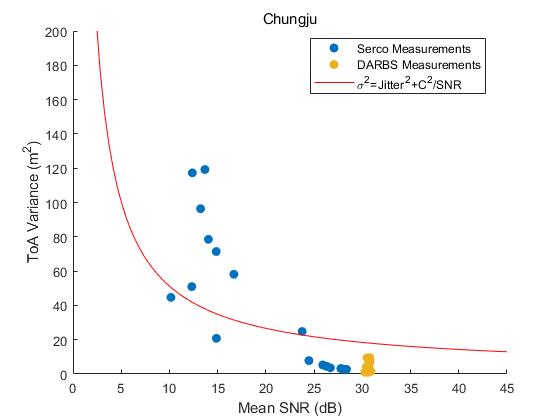}
    \caption{Model fit for Chungju transmitter. Yellow points represent measurements from the DARBS receiver, blue points represent measurements from the Serco receiver, and the red curve represents (\ref{eqn:MF_R-Mode_TOA_variance}).}
    \label{fig:Chungju}
\end{figure}

\begin{figure}
    \centering
    \includegraphics[width=0.8\linewidth]{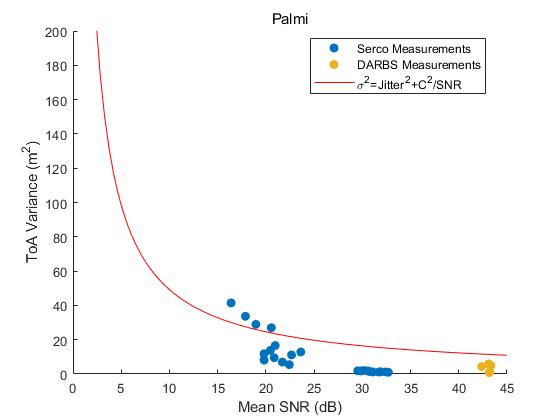}
    \caption{Model fit for Palmi transmitter. Yellow points represent measurements from the DARBS receiver, blue points represent measurements from the Serco receiver, and the red curve represents (\ref{eqn:MF_R-Mode_TOA_variance}).}
    \label{fig:Palmi}
\end{figure}

Fig.~\ref{fig:Eocheong}, Fig.~\ref{fig:Chungju} and Fig.~\ref{fig:Palmi} show the fit of the model to the actual measurements for Eocheong, Chungju and Palmi transmitters. 
The yellow points denote the empirical data collected using the DARBS receiver, while the blue points represent the data measured using the Serco receiver as used in \cite{Yu23:Empirical}. 
The x-axis shows $SNR_i$, and the y-axis displays the TOA variance ($\sigma^2_i$).
The model, represented by the red curve, is (\ref{eqn:MF_R-Mode_TOA_variance}).
The estimated values are as follows: the jitter value for Palmi is 0.00, for Chungju is 1.41, for Eocheong is 0.00, and the $C$ value is 22.15. 
Based on these values, $\sigma^2_i$ can be predicted for each transmitter's $SNR_i$.

\subsection{MF R-Mode Coverage Map}

\begin{figure}
\centering
    \includegraphics[width=1.0\linewidth]{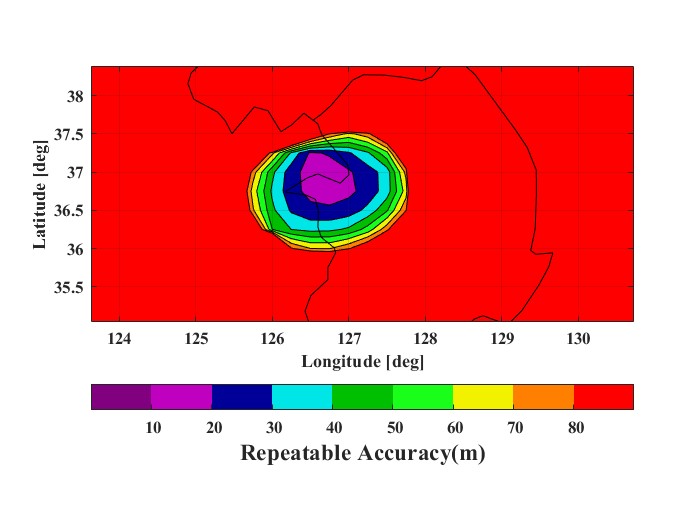}
\caption{Coverage map of the MF R-Mode system.}
\label{fig:coverage_map}
\end{figure}

Table~\ref{tab:parameters} lists the input parameters utilized in the simulation. 
Using the specified input parameters, the simulation generated the coverage map presented in Fig.~\ref{fig:coverage_map}.
This map visually represents the coverage calculated based on the transmitter locations and the signal strength at each location.

\begin{table}
\centering
\caption{Simulation Input Parameters}
\begin{tabular}{|l|c|}
\hline
\textbf{Simulation Input Parameters} & \textbf{Settings} \\
\hline
Eocheong Transmitter Power      & 300W \\
Palmi Transmitter Power         & 300W \\
Chungju Transmitter Power       & 500W \\
Eocheong Transmitter Jitter     & 0m \\
Palmi Transmitter Jitter        & 0m \\
Chungju Transmitter Jitter      & 1.41m \\
Season                          & 'Averaged' \\
Noise Level                     & 95\% \\
SNR Threshold                   & -15dB \\
\hline
\end{tabular}
\label{tab:parameters}
\end{table}

\section{Conclusion}
This study has enhanced the modeling of MF R-Mode TOA variance by incorporating additional data from the Eocheong transmitter alongside the Palmi and Chungju transmitters. 
By re-estimating jitter parameters for all three transmitters, we improved the system's repeatability accuracy and generated a comprehensive coverage map. 
A constraint of this research is its inability to fully consider time-variant factors such as changes in atmospheric noise.
This is because different receivers experience varying levels of internal noise and receive signals at different times. 
Future research should address this limitation to improve the accuracy of the model.
These efforts advance the simulation tools and contribute to the ongoing development of the R-Mode system in Korea.

\bibliographystyle{IEEEtran}
\bibliography{mybibfile, IUS_publications}

\vspace{12pt}

\end{document}